# 10 Inventions on Keyboard key layout
## A TRIZ based analysis

**Umakant Mishra**

Bangalore, India
umakant@trizsite.tk
http://umakant.trizsite.tk

**Contents**

## 1. Introduction

A keyboard is the most important input device for a standard computer. Today's keyboard is an evolution of a primitive keyboard through hundreds of inventions. This article illustrates 10 inventions on key layout of a computer keyboard.

**The key layout in a keyboard**

The keyboard consists of a set of keys, a key pressing mechanism and a connection the computer. The standard keyboards consists of some LEDs for indicating status of caps lock, num lock etc. According to the type of keys, there are four sections on the keyboard.

- A text entry section
- Navigation section
- Numeric keypad section, and,
- Function key section

Text entry section contains the standard character keys, navigation section contains cursor movement and page control keys, numeric keypad contains numeric keys and function keys section contain function keys and special keys.

**The character organization in a conventional keyboard**

The standard QWERTY keyboard was developed in the late 1800's for the typewriters. As people were acquainted with that the same layout was retained for the computer keyboards. Many people feel that the QWERTY layout is not very efficient layout and there have been many inventions on different layouts of character keys.

**Problems faced with the conventional keyboard**

Many researchers feel that the conventional QWERTY keyboard was designed for the mechanical typewriters and is not efficient to be used with a modern computer. While developing the keyboard for a mechanical typewriter, the purpose was to slow down the typing speed in order to avoid piling up striking levers on one another. But a computer keyboard does not have the limitations of those mechanical problems. Hence there is a scope for using a better key layout in a computer keyboard.

The drawbacks of the standard QWERTY keyboard are felt as below.

- It slows down the speed of typing
- Requires more movement of fingers and causes fatigue with the typist.
- Increases frequency of errors
- Loads hands and fingers with disproportionate amount of work.

## 2. The ideal key organization

In order to improve the key arrangement, two major issues should be addressed.

- The improved key arrangement should offer significantly improved productivity.
- The training time for learning the improved key arrangement should be minimized.

In order to address the above, we may have to consider the following issues.

**Consideration of human anatomy**

When we analyze the anatomy of human finger, we find some of our fingers are stronger and more dynamic than others. If we distribute the load properly so that the stronger fingers do more key press, the dynamic fingers to more movements and so on we can yield maximum advantage of the key layout.

10 Inventions on Keyboard key layout, by Umakant Mishra             http://www.trizsite.tk

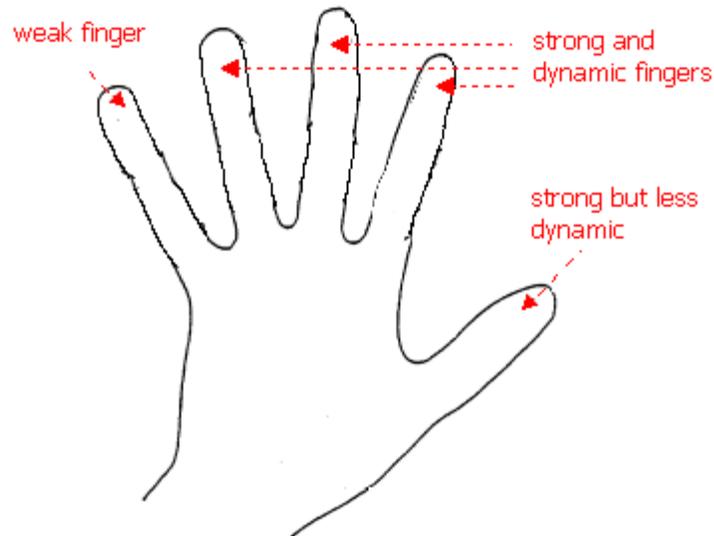

It is necessary to consider placing keys which have the most use on the home row, and of placing the most heavily used keys in position where they will be used by the strongest fingers.

**Consideration of home positioning**

On the other hand when we analyze positioning of the keys, the keys at the center are home positions for fingers. The fingers are most efficient when they are on the home positions. When the fingers move from their home positions they take time to reach to the desired key and they are likely to make mistakes.

Although it is well accepted that the fingers are fastest and least error prone if they are on their home positions, it is not possible to allocate 26 alphabet keys, 10 numeric keys and other special keys only to 8 or 10 home positions. That is why theoretically we need a layout which can maximize the use of home positions and minimize the movement of fingers.

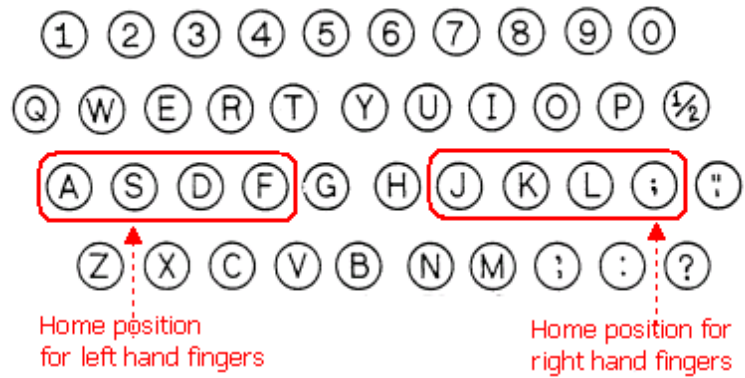

**Consideration of character frequencies**

There have been many studies to find the frequency of use of letters in English languages. The following table is a break up of different characters according to their frequency of use in 1000 characters.

```
E 130          L 36           W 16
T 92           H 34           V 15
N 79           C 31           B 10
R 76           F 28           X 5
O 75           P 27           Q 3
A 74           U 26           K 3
I 74           M 25           J 2
S 61           Y 19           Z 1
D 42           G 16
```

Although different types of documents may have different frequencies of usage and different studies might find some deviation in the above frequencies, by and large, all conclusions fall into a similar pattern.

This analysis is useful to refer while distributing workload among the fingers. While the more frequently used characters can be allocated to stronger fingers and home positions, the less frequently used characters can be allocated to the weak fingers or farther from home positions.

Besides, the figures should be considered to balance the workload between the two hands.

**Other considerations**
- The QWERTY keyboard has been in the market since long and people are so acquainted to it that there is a resistance for using any other layout even if it is better than QWERTY.
- Balance the workload between the left hand and the right hand so that the workload is more balanced between the two hands.
- Special layout may be useful in specific activities like number crunching, video gaming or any specific activity needing overuse of specific keys.
- Special layout required for very small keyboards required for laptops, palmtops and hand held computers.
- Special layout required for children, new learners, physically challenged users and other special users.

**The Ideal Final Result (IFR)**

The Ideal Final Result is achieved when the solution has all positive results and no negative results. The IFR in case of key layout can be as below.
- Working on keyboard should require no finger movements (or minimum movements)
- The layout should be such that user makes no error in typing (or minimum errors in typing).
- The keys should be easy to find and typing should be very easy to learn.
- Should not overburden hands or fingers.

## 3. Inventions on Key layout

### 3.1 Dvorak Keyboard: A new layout (Patent 2040248)

**Background**

Although QWERTY keyboard has been ruling over the keyboard kingdom since its invention by Sholes in 1868, many people feel that the distribution of the characters on a QWERTY keyboard is not suitable for a computer keyboard. As characters are scattered around the keyboard, it requires more finger movements and results in users' fatigue and typological errors. There is a need to solve this problem.

**Solution provided by the invention**

August Dvorak invented a different layout of keyboard based on scientific placement of characters to reduce the finger movements. The keyboard was later called Dvorak keyboard. The organization of the characters on a Dvorak keyboard is entirely different from the conventional QWERTY keyboard. The invention was patented in 1936 (US Patent 2040248).

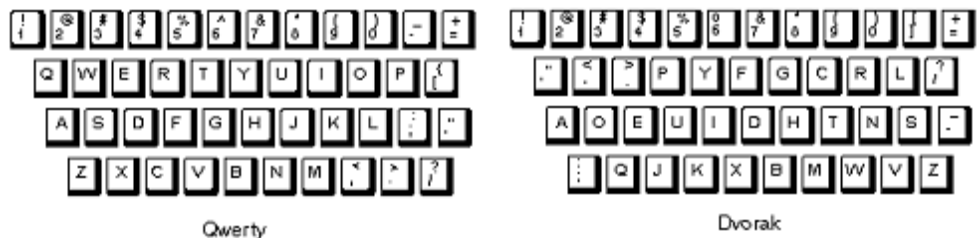

The Dvorak keyboard was proved to be efficient in terms of increasing the typing speed, decreasing typing errors, lessening fatigue of the typist and less cross movements of the fingers. Some companies tried to make Dvorak keyboards at different times, although it never got popularity and universal acceptance.

**TRIZ based analysis**

The invention uses a different layout of the keyboard **(Principle-17: Another Dimension).**

### 3.2 Keyboard with sequential key arrangement (Patent 4615629)

**Background problem**

The keys in a conventional QWERTY keyboard are scattered so randomly that it is difficult to locate any particular key on the keyboard. This is a great concern particularly for the beginners as they take long time for locating the keys and get frustrated or develop computer phobia. There is need for a keyboard that is very easy to learn even for a new user.

**Solution provided by the invention**

Daniel Power invented a keyboard (patent 4615629, Issued in Oct 86) with a simple organization of character keys. In this invention the keys are grouped in a unique alphabetical pattern, which avoids the difficulty of finding a key on the keyboard.

This keyboard is arranged in a vertical layout. The alphabetic keys are arranged in an alphabetical order (from A to Z) in nine rows with three keys per row.

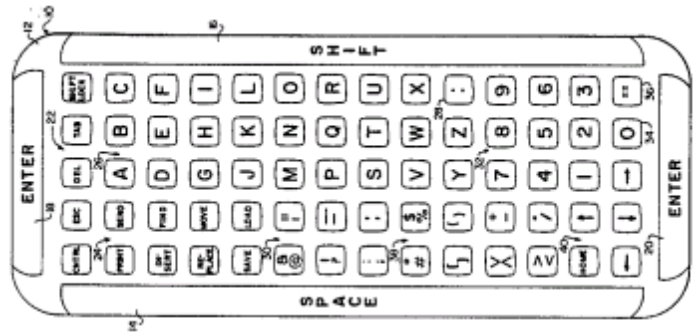

The three columns of character keys are to be effectively operated by the three stronger fingers of any single hand. There are two enter keys, one at the top and one at bottom. This alphabetic arrangement of key is suitable for beginners and non-typists.

**TRIZ based analysis**

The user need not learn the keyboard layout. Ideally, the layout should already be known to the user **(IFR)**.

The keys should be arranged in such a way that keys should be easy to find **(desired result)**.

The invention uses the normal alphabetical sequence of the keys, which is already known to any user **(Principle-25: Self service)**.

**3.3 Key arrangements and methods of use (Patent 5166669)**

**Background problem**

The standard QWERTY keyboard slows down the speed of typing because of excessive finger movements. The layout is complicated to cause more typological errors, loads the hands and fingers with disproportionate amounts of work and produces fatigue in the hands and figures of the typist.

Thus there is a need for an improved key arrangement for keyboard operators, which can give improved productivity and acceptable training time.

**Solution provided by the invention**

Roberg invented a keyboard called ASER D HN TIO (Patent 5166669, issued Nov 1992) with an improved key arrangement which provides better typing speed while maintaining many of the keys in the same position as the Qwerty key arrangement. By retaining many of the keys in the same position as in Qwerty, the new layout is supposed to be easy to learn for the existing users and requiring minimum training fro the new users. As the key locations are based on the frequency of characters and home position of the fingers, the keyboard needs less finger movements and is faster to type.

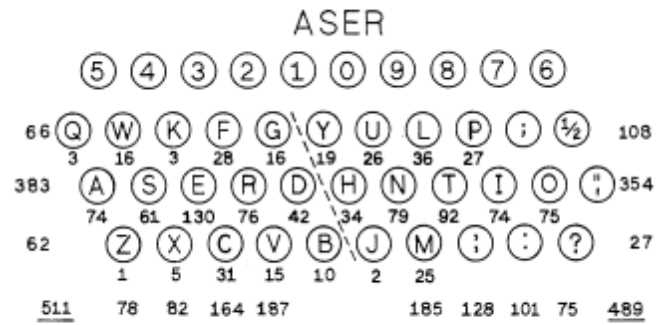

FIG. 4A

The invention also includes a software for converting existing QWERTY keyboards into the new key arrangement so that the existing users can continue using their old keyboards while the layout is changed internally by the software. This keyboard is named as ASER D HN TIO keyboard based on the keys at the home position.

**TRIZ based analysis**

A scientifically designed keyboard should follow certain rules like balancing the load on both the hands, reducing the length of finger movements, assigning the most repeating keys on the strongest fingers etc. If we keep these considerations in mind the key layout will vary a lot from the conventional QWERTY layout. This situation creates a contradiction. We should change the QWERTY layout, as it is not efficient, at the same time we should retain the QWERTY layout as it is most accepted **(Contradiction)**.

The new key arrangement is a balance between efficiency and QWERTY-like. It requires lesser hand movements while having substantial similarity with the conventional QWERTY keyboard for easy learning by existing operators **(Principle-17: Partial or excessive action).**

The invention includes a software which converts the strokes of the existing keyboards to the signals of the new layout so that the existing keyboards can still be used by just changing the key labels on the keytop. **(Principle-24: Intermediary, Principle-36: Conversion).**

### 3.4 Children's computer keyboard (Patent 5452960)

**Background problem**

The traditional keyboard is not designed for the minority group of users such as children or physically disabled individuals. Both these groups of users may have problem in locating the keys and moving their fingers precisely on the desired keys.

Jerry Wagstrom of Huntersville, N.C., developed a Kid Keys keyboard for children. The Kid Keys has oversized, colorful keys arranged in alphabetical order. The color arrangement in Kid Keys is that vowel keys are yellow, "R" key is red, "Y" key is yellow, "B" key is blue, and "G" key is green, and the rest of the keys are grey. However this has a limited benefit.

**Solution provided by the invention**

Kuhlenschmidt invented a children's computer keyboard (Patent 5452960, assignee- Nil, issued- Sep 1995), which includes enlarged keys that are color-coded according to their functions, and a four directional arrow-key pad of unique configuration. Characters on the keys of the children's computer keyboard are also enlarged for improved visibility. The enlarged keys provide wider top surface dimension between keys which provides each key with a larger error free area than a standard sized keyboard thus more tolerant for human error. A keyboard connector/extension cord with separable parts is provided for easy keyboard changing.

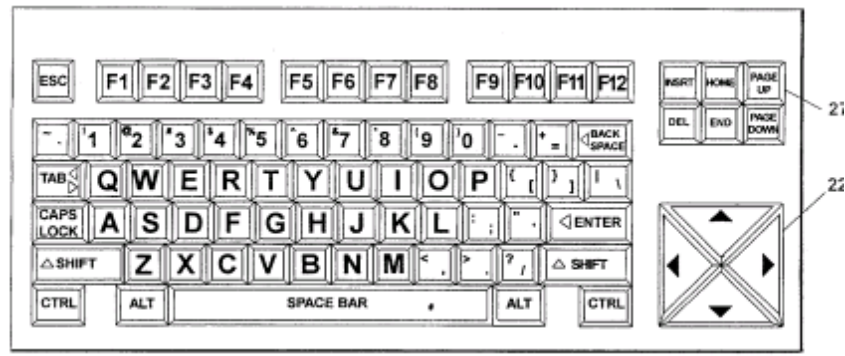

**TRIZ based analysis**

The keyboard uses enlarged keys and enlarged characters on the keys **(Principle-37: Expansion).**

The keys are color-coded according to their function. The alphabetic keys have the same color, the number keys have another color, the function keys have a third color, the arrow keys have a fourth color, punctuation mark keys have a fifth color etc. **(Principle-32: Color Change).**

The arrow keypad is square shaped which is easy to operate **(Principle-17: Another dimension).**

## 3.5 Computer keyboard adapter providing large size key surfaces (Patent 5514855)

**Background problem**

There are certain educational programs available for young children which do not use most of the keys in a 101 key conventional keyboard. The large number of additional keys on the keyboard creates confusion and leads to incorrect response. There is a need to have a special keyboard for educational software for young children.

**Solution provided by the invention**

Sullivan developed an adapter for the keyboard (Patent 5514855, assignee- Alpha Logic Inc, Issued May 1996), which amplifies specific keys of the keyboard to facilitate use by young children. The new keyboard has small number of large keys which when depressed, will cause depression of one or more keys of a selected area. There will be a software application which will associate the depressed key in the existing keyboard to determine the depressed key o the new keyboard.

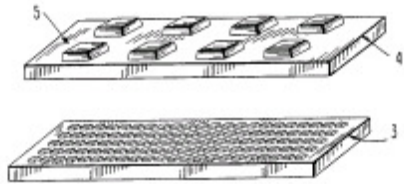

This new keyboard is intended to be placed on top of the existing keyboard so that other keys in the keyboard are protected from being depressed.

**TRIZ based analysis**

One solution is to protect all the keys on the keyboard with a thin hard cover except the specific keys that are required for the educational software **(Principle-2: Taking out)**. Change the labels or color of the specific key, which are available to be used with the software **(Principle-32: Color change)**. However, this solution does not reposition the required keys in an organized way for convenient access.

The current invention provides an expanded interface for specific required keys **(Principle-37: Expansion)**.

The invention keeps the new keyboard on top of the existing keyboard **(Principle-7: Nested doll)**.

The invention uses a software to determine the pressed switch in the new keyboard from the pressed keys of the existing keyboard **(Principle-36: Conversion)**.

### 3.6 Computer keyboard layout (Patent 5584588)

**Background problem**

The standard QWERTY layout was developed over 100 years ago for manual typewriter. The layout has neither any particular logic nor any particular sequence. Compared to operating a calculator where the fingers move either straight up (away from the user) or straight down (towards the user), the QWERTY keyboard requires various diagonal movements of fingers.

There is a need for a more logical keyboard layout which will make learning easier and speed up the work.

**Solution provided by the invention**

This invention was made by Gary Harbaugh (patent 5584588, issued Dec 1996) where the keys are arranged in alphabetical order (A-Z) which is easy to learn. Besides the alphabetic keys are arranged in straight rows and columns so that the fingers need to move only up and down avoiding the clumsy diagonal movements of a Qwerty layout.

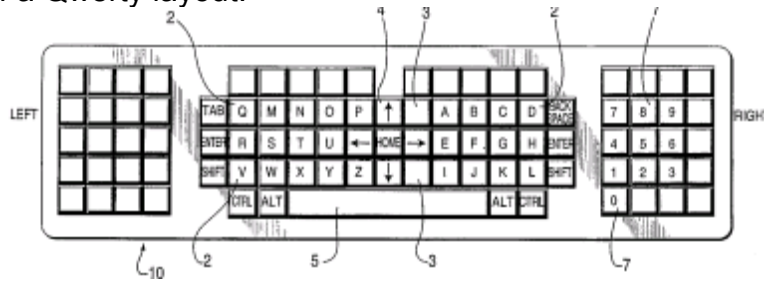

**TRIZ based analysis**

The invention uses a alphabetical sequence of the keys which is easy to learn, rather already known to everybody **(Principle-25: Self Service)**.

The keys are arranged in straight lines instead of diagonal arrangement of QWERTY layout **(Principle-17: Another dimension)**.

### 3.7 Enhancement of a QWERTY keyboard (Patent 5836705)

**Background**

The QWERTY keyboard, although popular and universally accepted, is not found to be suitable for faster typing. Although there are many inventions on scientific and better key layouts, none of them have been really accepted by people because of their wide variation from the popular QWERTY keyboard.

It is necessary to find a layout which should not be different from QWERTY but should be faster and efficient.

### Solution provided by the invention

There was an invention by John Choate (patent 5836705, Issued in Nov 98), which disclosed a method of a different key arrangement. According to the invention the home row has atleast three of the eight most used letters of the alphabet, The upper row has atleast three of the thirteen least used letters and the bottom row has at least for of the thirteen least used letters of the keyboard. The keyboard has atleast four and less than 26 of the keys have the same location as on the QWERTY keyboard.

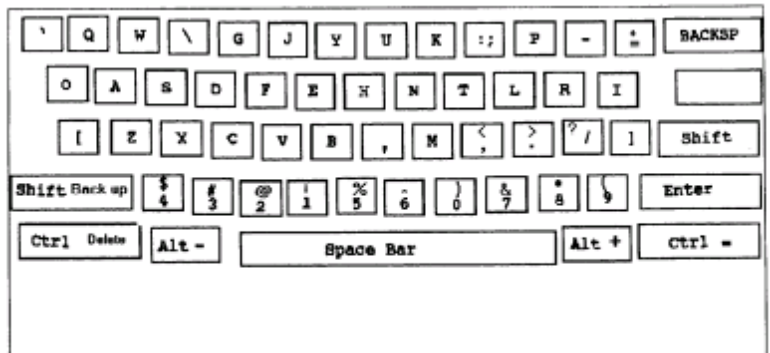

As there is a lot of similarity with QWERTY layout, the keyboard is easy to learn for the existing users. The new key layout reduces the wrist and elbow movements and reduces Repetitive Stress Injuries.

### TRIZ based analysis

Many faster keyboards in past have not been accepted by users because of their difference from the QWERTY layout. Hence, the keyboard should have a non-QWERTY layout to overcome the limitations of QWERTY. But the keyboard should be popular and acceptable as QWERTY. **(Contradiction)**

The invention discloses a layout, which has enough similarity with QWERTY layout for quick acceptance and popularity **(Principle-16: Partial or Excessive Actions).**

At the same time the layout is essentially different to provide speed and efficiency **(Principle-17: Another Dimension)**.

### 3.8 Keyboard having efficient layout of keys (Patent 6241406)

### Background problem

The conventional keyboard has some drawbacks. For example the letter keys are too concentrated, there are two sets of numeric keys and the locations of the two sets are not in equilibrium, and the function keys and symbolic keys are also irregularly arranged. Therefore, it is not easy for an operator to remember the locations of the keys, which leads to typing errors.

**Solution provided by the invention**

Yan invented a new layout of the keyboard (Patent 6241406, Jun 2001) with different key arrangements. The keyboard comprises a set of numerical keys arranged in the middle portion of the keyboard for inputting numerical information. The letter keys are arranged in the left and right side separated by the numeric keys. The location of the symbolic keys have been changed so as to conform with the left-right equilibrium characteristics of human brain, facilitate operation of the keyboard itself and save space.

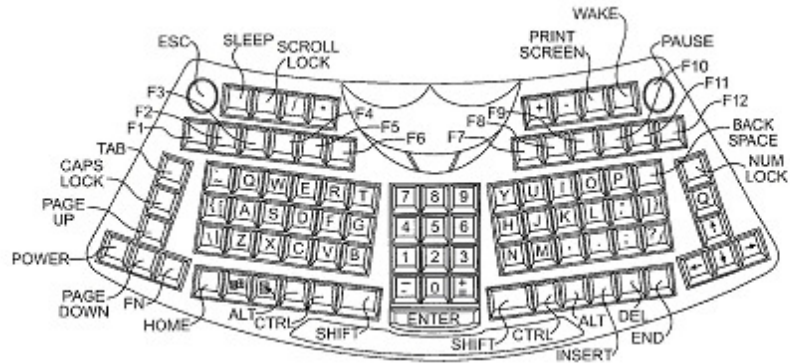

The keyboard of the invention is an improvement on the conventional keyboard in respect of the arrangement of the keys, it can speed up input, decrease the typing errors, and reduce the size of the keyboard.

**TRIZ based analysis**

The invention removes the extra set of numeric keys **(Principle-2: Taking out)**.

The invention rearranges the location of function keys, numeric and other keys of a conventional keyboard **(Principle-17: Another Dimension)**.

### 3.9 Keyboard and computer  (Patent 6398437)

**Background problem**

To input information from the keyboard quickly, the user should be allowed to concentrate on the display of the computer and documents. To do so, touch typing on blind typing in which the user types desired keys without looking at the keyboard is effecting. Cursor keys of a notebook personal computer typically are to the right of the character to be depressed with the little finger of the right hand. However, because the little finger is typically not strong enough to exert the required force, sometimes it is uneasy to operate the cursor keys with little finger. Therefore, the user often moves his right hand from the home position to operate the cursor keys, which disturbs the continuity of touch-typing and thereby affects productivity.

**Solution provided by the invention**

Yamazaki et al. invented a keyboard (patent 6398437, assigned to IBM, June 02) to eliminate these difficulties and facilitate a continuous touch-typing. According to the invention the cursor keys are disposed adjacently to the palm rest. The region is slanted to make distinguishable by touch from the main palm rest. The slanted region is located by tactile sensation. The keys are easily accessible without moving the hand, which makes the keyboard operable even in dark.

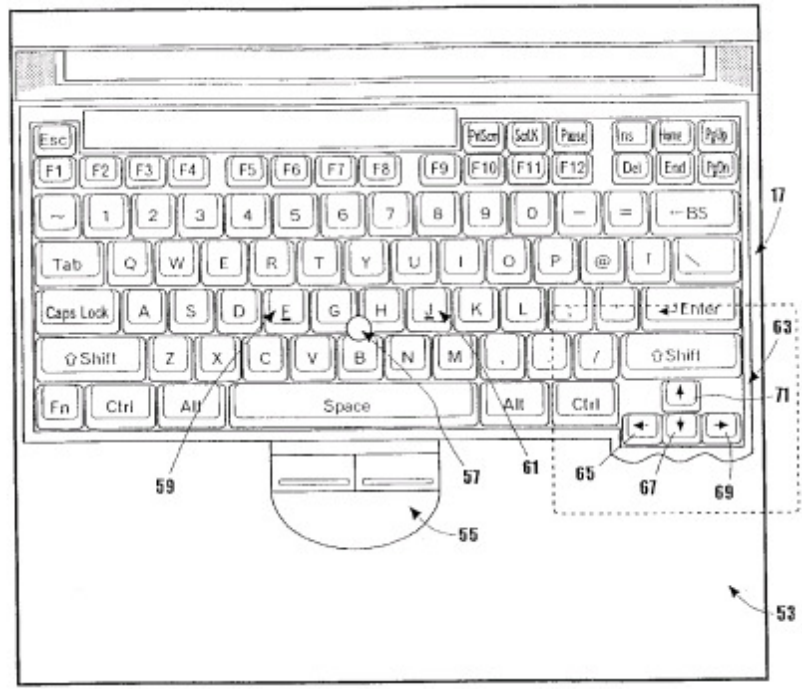

**TRIZ based analysis**

It should be possible to use cursor control keys without moving the arm from its home position and without moving your vision from the screen **(Ideal Final Result).**

The invention relocates the cursor keys close to the character keys, which can be accessed without moving the arm from home position **(Principle-17: Another Dimension)**.

The region having the cursor keys is slanted to make distinguishable by touch from the main palm rest. **(Principle-4: Asymmetry)**

### 3.10 Keyboard having buttons positioned for operation by heel of hand (Patent 6614421)

**Background problem**

The buttons on a keyboard are typically operated by the fingers or the thumb. The edge of the hands, although stronger than fingers, are not used in keyboard operation. How to use the heel of the hand to effectively operate the keyboard?

**Solution provided by the invention**

Selker et al. invented a keyboard (Patent 6614421, assigned to IBM, Sep 03) with built in pointing device, and left- and right-pointer control buttons operable by the thumbs. According to the invention the notebook computer includes additional left and right-pointer control buttons located to each side for easy operation by the outer edge of the hands.

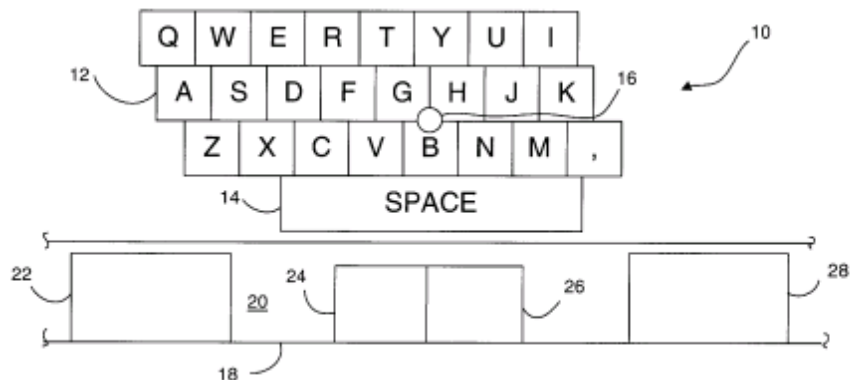

These additional control buttons duplicate the functions of the thumb operable buttons and are located and shaped such that they are natural and easy to use while typing and while using the finger tips to control the built in keyboard positioning device.

**TRIZ based analysis**

The invention positions some control buttons at the edge of the keyboard, which can be operated easily by the heel of the hand **(Principle-17: Another Dimension)**.

## 4. Summary and conclusion

**Reasons for changing the key layout**

As we saw there have been many attempts to change the conventional layout to many new layouts. The reason for changing the key layout on a keyboard can be one or more of the following.

- To reduce finger movements during typing.
- To achieve speed in typing.
- Reduce errors in typing
- Making the keyboard easy to learn

- Making easy for children to find the keys
- Reduce stress in hands and fingers
- Using in special devices or for special purposes

**The success of different inventions**

Although there have been many attempts to change the conventional QWERTY layout to a more scientific and efficient layout, no attempt has really been very successful in their acceptance. Some new layouts are definitely better than the conventional layout but they could not be commercially adopted because of the overwhelming dominance of QWERTY keyboard.

The only layout which had some acceptance and which was commercially available for long time was the Dvorak keyboard. Even now there is use and availability of Dvorak keyboards.

However, some special purpose keyboards have been implemented with special key layout, for example, in palmtops, in medical equipments etc. But the general-purpose keyboards are by-and-large found to be overwhelmingly dominated by the QWERTY layout.

**Conclusion**

The issue of key layout on a computer keyboard is very special. The conventional QUERTY layout is so popularly accepted that there is hardly any scope for accepting a new layout even if it is much better, simpler and efficient than the conventional one.
This experience of non-acceptance will, in one hand, reduce further inventions in this field, whereas on the other hand will encourage people to invent even better keyboards which can supercede the popularity of the conventional layout.